# Observation of dielectrically confined excitons in ultrathin GaN nanowires up to room temperature

Johannes K. Zettler,[1, a)] Pierre Corfdir,[1, b)] Christian Hauswald,[1] Esperanza Luna,[1] Uwe Jahn,[1] Timur Flissikowski,[1] Emanuel Schmidt,[2] Carsten Ronning,[2] Achim Trampert,[1] Lutz Geelhaar,[1] Holger T. Grahn,[1] Oliver Brandt,[1] and Sergio Fernández-Garrido[1]

[1)]*Paul-Drude-Institut für Festkörperelektronik, Hausvogteiplatz 5–7, 10117 Berlin, Germany*

[2)]*Institut für Festkörperphysik, Friedrich-Schiller-Universität Jena, Max-Wien-Platz 1, 07743 Jena, Germany*

The realization of semiconductor structures with stable excitons at room temperature is crucial for the development of excitonics and polaritonics. Quantum confinement has commonly been employed for enhancing excitonic effects in semiconductor heterostructures. Dielectric confinement, which is potentially much stronger, has proven to be more difficult to achieve because of the rapid nonradiative surface/interface recombination in hybrid dielectric-semiconductor structures. Here, we demonstrate intense excitonic emission from bare GaN nanowires with diameters down to 6 nm. The large dielectric mismatch between the nanowires and vacuum greatly enhances the Coulomb interaction, with the thinnest nanowires showing the strongest dielectric confinement and the highest radiative efficiency at room temperature. In situ monitoring of the fabrication of these structures allows one to accurately control the degree of dielectric enhancement. These ultrathin nanowires may constitute the basis for the fabrication of advanced low-dimensional structures with an unprecedented degree of confinement.

---

[a)]equal contribution

[b)]Electronic mail: corfdir@pdi-berlin.de; equal contribution

In 1979, Keldysh demonstrated that excitonic phenomena in semiconductors can be dramatically enhanced by combining materials with significantly different dielectric constants (or, in modern terms, relative permittivities).[1] If the dielectric constant of the semiconductor is larger than that of the surrounding material, the redistribution of the polarization field caused by the dielectric mismatch at the interfaces results in an enhancement of the Coulomb interaction between free electrons and free holes. This effect, denoted as the dielectric confinement of the exciton, gives rise to a greatly increased exciton binding energy and oscillator strength[2–5] and is thus interesting for room temperature optoelectronic applications involving excitons[6–10] such as polariton lasers[6] or excitonic switches.[7]

The strongest dielectric confinement possible is realized in semiconductor nanostructures surrounded by a material with a dielectric constant of 1, i. e., vacuum or air.[11] However, the unpassivated surfaces of semiconductors are believed to invariably suffer from efficient surface recombination, rendering such structures unsuitable for optoelectronic devices. For instance, the surface recombination velocity at the sidewalls of GaAs nanowires is on the order of $5 \times 10^5$ cm/s.[12] As a result, GaAs nanowires with diameters below 100 nm do not emit any light at all.[13] Passivating the surface of GaAs nanowires with an (Al,Ga)As shell reduces the surface recombination velocity by orders of magnitude,[14] but at the same time also the dielectric mismatch. Consequently, the impact of the dielectric confinement on the exciton in semiconductor heterostructures is negligible.[15] ZnO nanowires exhibit intense excitonic emission at room temperature,[16] but the dielectric constant of ZnO is too small to give rise to a strong dielectric confinement. Stronger dielectric confinement of the exciton may, in principle, be obtained in hybrid structures combining insulating and semiconducting materials,[17] but the incompatible crystal structures lead inevitably to a high density of interface states and thus high interface recombination velocities.

In this context, GaN nanowires appear to be a more suitable platform for optoelectronic applications based on dielectrically enhanced excitons. GaN, the compound semiconductor that enables solid-state lighting,[18] is also a promising material for excitonics because the exciton binding energy is equal to $k_B T$ at room temperature already in the bulk. GaN nanowires form spontaneously on a wide variety of substrates and are free of homogeneous strain and threading dislocations.[19] In addition, the dangling bond states for GaN polar and nonpolar surfaces are located far from midgap[20] so that surface recombination is expected to be comparatively inefficient. In fact, a surface recombination velocity smaller than 210 cm/s was reported recently for unpassivated GaN nanowires,[21] a value three orders of magnitude smaller than that measured for GaAs nanowires[12] and comparable to the interface recombination velocity in planar GaAs/(Al,Ga)As double heterostructures grown by molecular beam epitaxy.[22]

Here, we report on room temperature emission from dense ensembles of ultrathin epitaxial GaN

nanowires with diameters as small as 6 nm. These ultrathin nanowires were fabricated by thinning as-grown nanowire ensembles using a post-growth thermal decomposition process taking place in vacuum. In contrast to GaAs nanowires,[23] thinned GaN nanowires show intense luminescence at room temperature even in the absence of surface passivation, demonstrating that the impact of surface recombination in this material system is indeed minor. The thinnest nanowires are in fact those exhibiting the largest radiative efficiency at elevated temperatures. We show that the mismatch in dielectric constants at the nanowire surface leads to a strong enhancement of excitonic effects, reducing the exciton's radiative lifetime at room temperature and promoting radiative processes over nonradiative ones. Since the fabrication of these thinned nanowires is monitored in situ, we can control the degree of dielectric confinement of the exciton.

Figure 1(a) presents a scanning electron micrograph of a typical GaN nanowire ensemble grown by molecular beam epitaxy on a Si substrate (see Methods). These nanowires exhibit a length of 1.6 μm and an areal density of $9 \times 10^9$ cm$^{-2}$. The 51 nm diameter of these nanowires is much larger than the exciton Bohr radius in GaN ($a_B = 3$ nm), and neither dielectric nor quantum confinement is expected. Dielectric confinement of the exciton starts to be noticeable for diameters of $5a_B$.[5,24,25] Nanowires this thin are difficult to obtain by direct growth.[19] For reducing the diameter of as-grown GaN nanowires, we used a post-growth decomposition process similar to the one described in Refs. 23 and 26 (see Methods). As shown in Fig. 1(b), annealing an as-grown nanowire ensemble for 30 min at 920 °C in ultra-high vacuum (UHV) results in a reduction in the length and diameter of the nanowires. The analysis of cross-sectional and top-view micrographs reveals that the average length decreases to approximately 1.1 μm. Although partially decomposed nanowires also appear to be much thinner than the as-grown ones, a reliable assessment of the final nanowire diameter by scanning electron microscopy is not straightforward. Under electron irradiation, the nanowires bend and bundle due to charging, leading to an overestimation of their diameter when analyzing top-view and cross-sectional images.[23,27] Upon thermal decomposition, the nanowires also exhibit a pronounced tapering, suggesting that the nanowire diameter decreases faster at their tip than at the bottom part. For instance, the average base diameter for the nanowires in Fig. 1(b) is 32 nm, while the diameter along the top 600 nm of the nanowire in Fig. 1(c) decreases from approximately 20 to 6 nm. Our systematic observation of wire tapering contrasts with the homogeneous reduction in nanowire diameter during thermal decomposition reported in Ref. 26. This discrepancy might be related to the fact that in contrast to Ref. 26, where the decomposition process took place in $H_2$ or $NH_3$ atmospheres, we decomposed our nanowires in UHV, making it possible to carry out this process directly in the molecular beam epitaxy chanber, and to control it in situ by quadrupole mass spectrometry

(QMS).

High-resolution transmission electron micrographs taken close to the tip of individual partially decomposed nanowires are shown in Figs. 1(d) and 1(e). The nanowires are single crystals and are free of any dislocations as the as-grown nanowires.[28] Despite the pronounced tapering detected by scanning electron microscopy [Fig. 1(c)], the nanowire sidewalls remain atomically abrupt upon thermal decomposition. The direct observation of atomic planes allows one to accurately determine the nanowire diameter, which we find equal to $(8.8 \pm 0.3)$ and $(5.8 \pm 0.39)$ nm in Figs. 1(d) and 1(e), respectively. Figure 1(f), taken at the top of the nanowire shown in Fig. 1(d), shows that the tip diameter of this thinned nanowire is below 5 nm.

As shown above, thermal decomposition facilitates the fabrication of ultrathin GaN nanowires while preserving their high structural perfection. In contrast to previous decomposition studies,[23,26] we monitor the decomposition in situ by QMS[29–31] and can thus control this process. Since the dissociation of GaN in UHV is thermally activated,[32–34] the decomposition rate of GaN nanowires is controlled primarily by the substrate temperature. Figure 2(a) depicts the change of substrate temperature and the resulting temporal evolution of the desorbing Ga flux $\Phi_{G}^{des}$ as measured by QMS in units of ion current during the post-growth decomposition of the sample shown in Fig. 1(b). For the decomposition, the temperature was ramped within 10 min from 680 to 920 °C and was then kept constant for 30 min. Note that, after reaching 920 °C, the reading of the substrate temperature (acquired with an optical pyrometer) was not constant, but steadily decreased until stabilizing at 905 °C. This effect is attributed not to actual temperature changes but to variations in the optical emissivity of the sample. The temperature dependence of $\Phi_{G}^{des}$ displayed in Fig. 2(b) has been obtained from the 10 min temperature increase in Fig. 2(a) (see the methods section for more details on the conversion of $\Phi_{G}^{des}$ to equivalent growth rate units of nm/min). Within this temperature range, $\Phi_{G}^{des}$ increases by almost two orders of magnitude and is as high as 15 nm/min at 920 °C. Despite the thermally-induced morphological changes in the nanowire ensemble (Fig. 1), $\Phi_{G}^{des}$ approximately follows an Arrhenius-like temperature dependence [Fig. 2(b)]:

$$\Phi_{G}^{des}(T_S) = C_1 \exp[-E_D/(k_B T_S)] \, , \tag{1}$$

where $E_D$ denotes the activation energy for the nanowire decomposition process, $T_S$ the substrate temperature, and $C_1$ a constant. The best fit to the data yields $E_D = (3.1 \pm 0.1)$ eV, identical to the value reported for the layer-by-layer decomposition of GaN(0001) films in vacuum.[34]

In contrast to the layer-by-layer decomposition of GaN films, where $\Phi_{G}^{des}$ remains constant at a fixed temperature,[34] this quantity for a nanowire ensemble decays exponentially [Figs. 2(a) and 2(c)]. There-fore, in the nanowire case, the decomposition rate is proportional to the amount of remaining material. Extrapolating $\Phi_G^{des}$ to the background level allows one to estimate the total GaN volume fraction that has been decomposed during the thermal decomposition process. For instance, the QMS data in Fig. 2(a) indicate that a 30 min annealing at 920 °C in UHV leads to the decomposition of $(96 \pm 2)\%$ of the material. To confirm this analysis, we performed quantitative Rutherford backscattering measurements on this sample and on the as-grown nanowire ensemble, and we obtained a volume fraction of decomposed material of $(96.2 \pm 1.6)\%$ (see Supplementary Information). QMS appears to be sufficiently accurate for a quantitative in situ control of the decomposition of GaN nanowires, enabling the fabrication of ultrathin nanowire ensembles with tailored dimensions.

The temporal evolution of $\Phi_G^{des}$ gives further insight into the decomposition of GaN nanowires in UHV. The micrographs in Fig. 1 indicate that both the diameter and the length reduce over time. Based on the value of $E_D$ [Fig. 2(a)] and considering the smooth sidewalls after partial decomposition [Figs. 1(d)–1(f)], we infer that the top and side facets of the nanowire decompose layer-by-layer, as illustrated in Figs. 2(d) and 2(e). Assuming that the decomposition rate is limited by the creation of kinks at the edges between the sidewalls and the top facet, it is possible to obtain an analytic expression for the temporal evolution of $\Phi_G^{des}$ (see Supplementary Information). Despite its simplicity, our modeling describes the temporal evolution of $\Phi_G^{des}$ very well, as shown by the line in Fig. 2(c). We can thus deduce the decomposition rate constants for the nanowires side and top facets ($R_s$ and $R_t$, respectively), and the best fit to the data yields $R_s = 2.5 \times 10^{-4}\ \mathrm{s}^{-1}$ and $R_t = 2.8 \times 10^{-3}\ \mathrm{s}^{-1}$ (see Supplementary Information). The top facets decompose faster than the sidewalls, in agreement with our experimental observations (Fig. 1). Our understanding on the evolution of the shape of the nanowires during thermal decomposition is illustrated and summarized in Figs. 2(f)–2(h). Prior to decomposition, the nanowires have a (more or less regular) hexagonal cross-sectional shape with an abrupt vertical edge between side and top facets [Fig. 2(f)]. The decomposition of the facets proceeds layer-by-layer and is initiated at the edges between the side and top facets [Fig. 2(g)]. As explained in Supplementary Information, we assume for simplicity in our decomposition model that the creation of kinks at the edges between the side and top facets is the rate limiting step for the decomposition process. Consequently, the decomposition rate is proportional to the nanowire's cross-sectional perimeter at the top facet. Nevertheless, in reality, a new layer is likely to start decomposing before the previous one has reached the substrate, explaining why thinned GaN nanowires obtained by thermal decomposition are tapered, as illustrated in Fig. 2(h). Because of this tapering and the experimental difficulty to obtain the diameter of the nanowires at their tip, we will characterize the nanowires in all what follows by their base diameter $d_B$. This latter value can be reliably obtained by a statistical analysis of cross-sectional scanning electron micrographs and was found to scale with the decomposed volume as obtained by QMS. Note, however, that the nanowires are much longer than the diffusion length of excitons in GaN.[35] Therefore, their emission properties are not determined by their base diameter, but rather by their diameter closer to the tip. These values are very different: for instance, the thinned nanowires obtained after a 30 min annealing at 920 °C shown in Figs. 1(b)–1(f) exhibit a base diameter of $d_B = 32$ nm, but a diameter close to the tip of less than 10 nm [cf. Fig. 1(d)].

Even this latter value is significantly larger than the Bohr radius of the exciton in bulk GaN ($a_B = 3$ nm). Therefore, the impact of quantum confinement on the energy of the exciton is not significant in any of the samples under investigation. Yet, as shown in Fig. 3(a), the photoluminescence peak energy at 300 K for ensembles of partially decomposed GaN nanowires increases from 3.412 to 3.454 eV with progressive decomposition. This observation is a signature of dielectric confinement: the image charges resulting from the large mismatch between the relative dielectric constants of GaN and vacuum lead to a renormalization of the bandgap and to an increase in the strength of the Coulomb interaction.[1,3,4,11] With decreasing $d_B$, the radial dielectric confinement of the exciton becomes stronger, and consequently, the photoluminescence line of thinned nanowires is blueshifted. Furthermore, the smaller $d_B$, the larger the emission broadening [Fig. 3(a)], since the emission energy increasingly depends on diameter. A similar behavior is observed at 5 K, where the near band-edge emission of the GaN nanowire ensemble is dominated by the recombination of excitons bound to donors and stacking faults [Fig. 3(b)].[21] Both the donor-bound and the stacking fault-bound exciton transitions blueshift and broaden with decreasing nanowire diameter.

Since we are dealing with bare GaN nanowires with free, unpassivated surfaces, one would expect surface recombination to become more important for the thinner nanowires. The emission intensity from the ultrathin nanowires should thus be considerably weaker than for the as-grown ensemble, as reported for GaAs nanowires.[13,23] Figure 3(c) compares the photoluminescence intensity at 300 K of the as-grown nanowires with that of thinned nanowires with $d_B = 32$ nm. The photoluminescence spectrum of the thinned nanowires was normalized by the volume fraction of material left after decomposition. The normalization factor was obtained independently from the QMS analysis shown in Fig. 2(a) and from Rutherford backscattering experiments (Supplementary Information). In contrast to what is expected in the presence of efficient surface recombination, the normalized integrated emission intensity for the partially decomposed nanowires is actually larger (2.6 times) than that of the as-grown ensemble [Fig. 3(c)]. This high emission intensity from the ultrathin nanowires does not originate from an increase in nanowire-light coupling with decreasing diameter. In fact, in our experimental geometry (see Methods), the light coupling between the thinned nanowires and free space is reduced compared to the as-grown nanowires.[36] The observation in Fig. 3(c) thus demonstrates that nonradiative recombination at the sidewalls of partially decomposed nanowires at 300 K is negligible in comparison to other nonradiative recombination processes. In the remainder of this work, we show that the intense emission observed for the ultrathin nanowires is a direct consequence of the dielectric confinement of the exciton in these structures.

To get a deeper insight into the radiative efficiency of ultrathin GaN nanowires, we investigate our samples by time-resolved photoluminescence spectroscopy. For spontaneously formed GaN nanowires grown by molecular beam epitaxy, the exciton decay has been shown to be largely nonradiative even at low temperatures, as reflected by decay times $\tau_{PL}$ between 150 and 200 ps.[37] Figure 4(a) shows photoluminescence transients recorded at 5 K from the samples under investigation. The decay times derived from the photoluminescence transients are shown in the inset. For the as-grown ensemble, we observe a decay time of 140 ps, similar to those obtained from comparable samples studied previously.[21,37] Strikingly, the decay time of the thinned nanowires is not shorter, but (with the exception of the sample with the thinnest nanowires, where surface recombination may also come into play) longer than that of the as-grown ensemble. Since the decay of excitons in these as-grown GaN nanowires at 5 K is dominated by nonradiative recombination at point defects,[37] this observation suggests a reduction in point defect density after the thermal decomposition process.

Figure 4(b) shows the temperature dependence of the near band-edge emission intensity for the samples under investigation. The photoluminescence intensity for thick nanowires follows a $T^{-3/2}$ dependence. This behavior is characteristic for spontaneously formed GaN nanowires in general and reflects the increase in radiative lifetime with temperature with the nonradiative lifetime staying essentially constant.[37] Remarkably, the quenching in photoluminescence intensity with temperature for our thinned nanowires is less pronounced. In particular, for the ensemble with $d_B = 27$ nm, the photoluminescence intensity remains constant between 5 and 150 K and quenches only for higher temperatures. As for the as-grown nanowires, the quenching is mainly controlled by the temperature dependence of the radiative lifetime. For example, the photoluminescence intensity of the sample with a base diameter of 32 nm is reduced by a factor of 70 when increasing the temperature from 5 to 300 K. In the same temperature interval, the photoluminescence decay time $\tau_{PL}$ (which essentially represents the nonradiative lifetime) is reduced by a factor of only 3.3 as shown in Fig. 4(c).

Since relaxation processes are fast in comparison to recombination mechanisms, the radiative lifetime is proportional to the inverse of the emission intensity at zero delay (see Supplementary Information). Figure 4(d) shows the thus obtained evolution of the radiative lifetime as a function of temperature for all samples normalized to the value of the radiative lifetime at 5 K ($\tau_{r,5K}$). The thinner the nanowires, the slower the increase in radiative lifetime between 5 and 300 K [Fig. 4(d)], and thus the weaker the quenching of the photoluminescence intensity [Fig. 4(b)].

Due to the delocalization of carriers in **k**-space with increasing temperature, the radiative lifetime of a population of excitons with a dimensionality $n$ increases as $T^{n/2}$.[21] The slower decrease in emission intensity observed for the thinnest nanowires in Fig. 4(b) is thus evidence of reduced exciton dimensionality in these structures. In other words, the impact of the dielectric mismatch in ultrathin GaN nanowires is strong enough to quantize the center-of-mass motion of the exciton perpendicular to the nanowire axis, i. e., partially decomposed GaN nanowires are actually quantum wires.

Note that below 200 K, the decrease in emission intensity for our thinnest nanowires is even smaller than the $T^{1/2}$-dependence expected for ideal quantum wires [Fig. 4(b)].[38] We attribute this finding to the fact that we do not deal with a purely one-dimensional exciton distribution, but rather with one-dimensional free excitons thermalized with a distribution of zero-dimensional excitons bound to donors and to stacking faults [Fig. 3(b)]. The presence of these localized states and their coupling with free excitons[21,37] leads to a smaller increase in exciton radiative lifetime with temperature.[35,39] Finally, the delocalization of excitons from donors and stacking faults leads to an increase in exciton coherence length, giving rise to the decrease in radiative lifetime observed between 50 and 150 K for thinned nanowires with $d_B = 27$ nm [Fig. 4(d)].

In conclusion, we have reported on the fabrication of ultrathin GaN nanowires. These unpassivated nanowires exhibit intense excitonic emission at room temperature, demonstrating that surface recombination at the sidewalls of GaN nanowires is slow in comparison to other recombination processes. Excitonic effects in thinned nanowires are enhanced by the mismatch in dielectric constants of GaN and vacuum. The dielectric confinement of the exciton gives rise to an increase in the internal radiative efficiency, which is largest for the thinnest nanowires. Thinned GaN nanowires are thus a promising system for optoelectronic devices operating in the strong coupling regime.[40,41] In particular, microcavities with thinned (In,Ga)N nanowires as an active medium could make it possible to achieve polariton lasing in the visible range. Ultrathin nanowires form also an attractive platform for quantum optics applications. For instance, the in situ control of the nanowire diameter facilitates the realization of quantum dots with well-defined lateral shape and thus light emission properties, and it paves the way for the tuning of the radial confinement of the exciton in crystal-phase quantum dots.[21,42]

## I. METHODS AND EXPERIMENTS

**Nanowire growth.** All samples were grown on Si(111) substrates in a molecular beam epitaxy system equipped with a solid-source effusion cell for Ga as well as a radio-frequency $N_2$ plasma source for active N. The impinging Ga and N fluxes were calibrated in GaN-equivalent growth rate units for planar layers of nm/min, as explained in Ref. 43. A growth rate of 1 nm/min corresponds to $7.3 \times 10^{13}$ atoms cm$^{-2}$ s$^{-1}$. The substrate temperature was measured using an optical pyrometer calibrated with the $1 \times 1$ to $7 \times 7$ surface reconstruction transition of Si(111) occurring at about 860 °C.[44] The as-received 2$^{\prime\prime}$ Si(111) substrates were etched using diluted (5%) HF. Prior to the growth, the substrates were outgassed at 920 °C for 30 min to remove any residual $Si_xO_y$ from the surface. After exposing the substrates for 10 min at growth temperature to an active N flux of $\Phi_N = (7.8 \pm 0.5)$ nm/min, nanowire growth was initiated by opening the Ga shutter. The samples were grown using a Ga flux of $\Phi_{Ga} = (4.5 \pm 0.5)$ nm/min and a N flux of $\Phi_N = (7.8 \pm 0.5)$ nm/min. The substrate temperature was kept at 800 °C during the first 25 min of the growth and then raised to 810 °C. As described in Ref. 45, this two-step growth procedure facilitates the fabrication of homogeneous nanowire ensembles with small diameters and low coalescence degrees without suffering from a long incubation time. The growth was terminated after a total growth time of 4 h by closing simultaneously the Ga and N shutters. As commonly reported, the resulting GaN nanowires crystallize in the wurzite crystal structure, are oriented along the *c*-axis, and exhibit flat $(000\bar{1})$ and $\{1\bar{1}00\}$ top and side facets, respectively.

**Nanowire decomposition.** After the growth of the GaN nanowire ensemble, the N-plasma was switched off to re-establish UHV conditions, and the substrate temperature was reduced to 600 °C. For the subsequent decomposition step, the substrate temperature was then raised to 920 °C. The desorbing Ga flux $\Phi_{Ga}^{des}$ was monitored in situ during the experiments by line-of-sight quadrupole mass spectrometry (QMS). The QMS response to the $Ga^{69}$ signal was also calibrated in GaN-equivalent growth rate units for planar layers of nm/min, as explained in detail in Refs. 29–31. The decomposition of the nanowire ensemble was terminated by cooling down the substrate.

**Electron microscopy.** Scanning electron microcoscopy (SEM) was performed using a microscope equipped with a field-effect gun operating with an acceleration voltage of 5 kV. The average base diameter $d_B$ was obtained for each sample by a statistical analysis of cross-sectional SEM images. Individual nanowires were mechanically dispersed on a Si substrate to study their shape after thermal decomposition. To analyze the microstructure of partially decomposed nanowires, high-resolution transmission electron microscopy (HRTEM) was performed on a large number of individual nanowires using a high resolution analytical electron microscope operating at 300 kV, equipped with a slow-scan charge-coupled device camera. Cross-sectional TEM specimens were prepared using conventional techniques, where the nanowires are glued for mechanical stabilization and thinned by mechanical grinding, dimpling, and Ar-ion milling.[28] Within a given sample, all investigated nanowires exhibited a similar shape and were found to be free of dislocations.

**Spectroscopic investigation.** The emission from ensembles of as-grown and partially decomposed GaN nanowires were probed by continuous-wave (cw) and time-resolved (TR) photoluminescence (PL) spectroscopy. We carried out cw-PL measurements with the 325 nm line of an HeCd laser. The diameter of the excitation spot was 60 μm, and the excitation density was kept below 10 mW/cm². The nanowire emission was dispersed by an 80 cm spectrograph and detected by a charge-coupled device camera with a linear detection efficiency. For TR PL experiments, the second harmonic of fs pulses obtained from an optical parametric oscillator pumped by a Ti:sapphire laser was used (excitation wavelength and repetition rate of 325 nm and 76 MHz, respectively). The energy fluence per pulse was kept below 0.3 μJ/cm². The transient emission was dispersed by a 22 cm spectrograph and detected by a streak camera that can be operated in synchroscan or in single shot mode. For both cw and TR PL, the samples were mounted in a coldfinger cryostat, whose temperature was varied between 5 and 300 K. For all experiments, the laser propagation direction and polarization axis were parallel and perpendicular to the nanowire axis, respectively.

## II.   ACKNOWLEDGMENTS

We thank Alberto Hernández-Mínguez for a careful reading of the manuscript, and Hans-Peter Schönherr and Carsten Stemmler for their dedicated maintenance of the molecular beam epitaxy system. We acknowledge partial financial support by the Deutsche Forschungsgemeinschaft within SFB 951.

## III.   AUTHOR CONTRIBUTIONS

J.K.Z. fabricated the samples and realized the decomposition experiments. P.C., C.H. and T.F. carried out the photoluminescence experiments. E.L. and A.T. acquired and analyzed transmission electron micrographs. U.J. carried out secondary electron microscopy. E.S. and C.R. performed the RBS experiments and analysis. J.K.Z., O.B. and S.F.G. analyzed the QMS data and modeled the decomposition process. P.C. and O.B. analyzed the spectroscopic data. J.K.Z., P.C., L.G., H.T.G., O.B. and S.F.G. wrote the paper. O.B. and S.F.G. planned and designed the experiments, with contributions of L.G. All authors contributed to the discussion of the results and revised the manuscript.

## IV.   COMPETING FINANCIAL INTERESTS

The authors declare no competing financial interests.

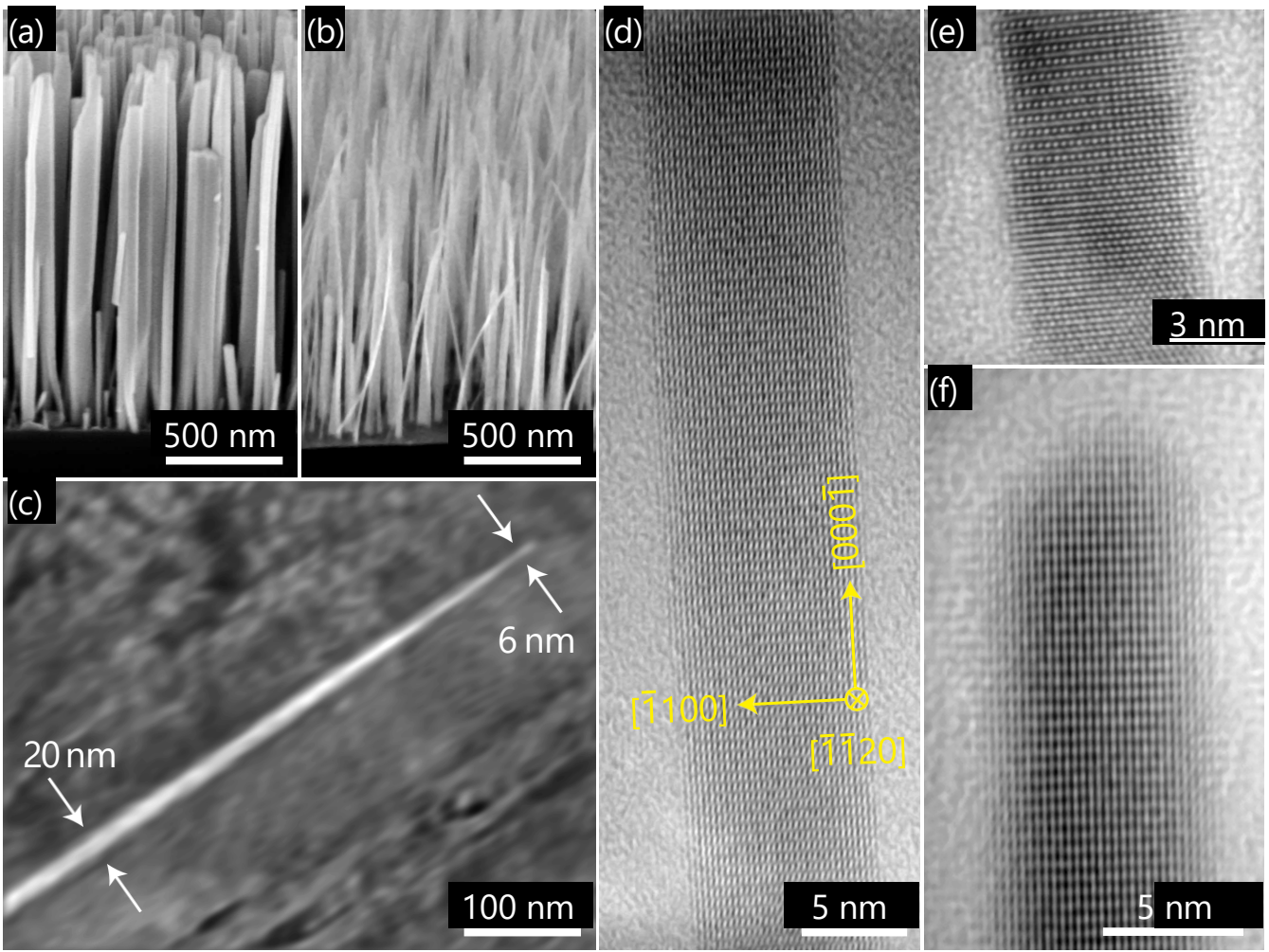

FIG. 1. **Sample morphology before and after thermal decomposition. a, b,** Bird's eye view scanning electron micrographs of GaN nanowire ensembles **a** before and **b** after thermal decomposition for 30 minutes at 920 °C in UHV. The nanowire length decreases from 1.6 to 1.1 μm. The areal density for the as-grown nanowires is $9 \times 10^9$ cm$^{-2}$. **c,** Scanning electron micrograph of a partially decomposed GaN nanowire dispersed on a Si substrate. The decomposition process results in some tapering. The diameter at the tip of the nanowire is 6 nm. **d, e,** High-resolution transmission electron micrographs reveal that the partially decomposed nanowires are single crystals and exhibit smooth sidewalls. **f,** High-resolution transmission electron micrograph of the tip of the nanowire shown in **d**. The micrographs **d–f** have been enhanced by a Fourier filter selecting the relevant spatial frequencies corresponding to the distance of the $\{0002\}$ and the $\{1\bar{1}00\}$ planes.

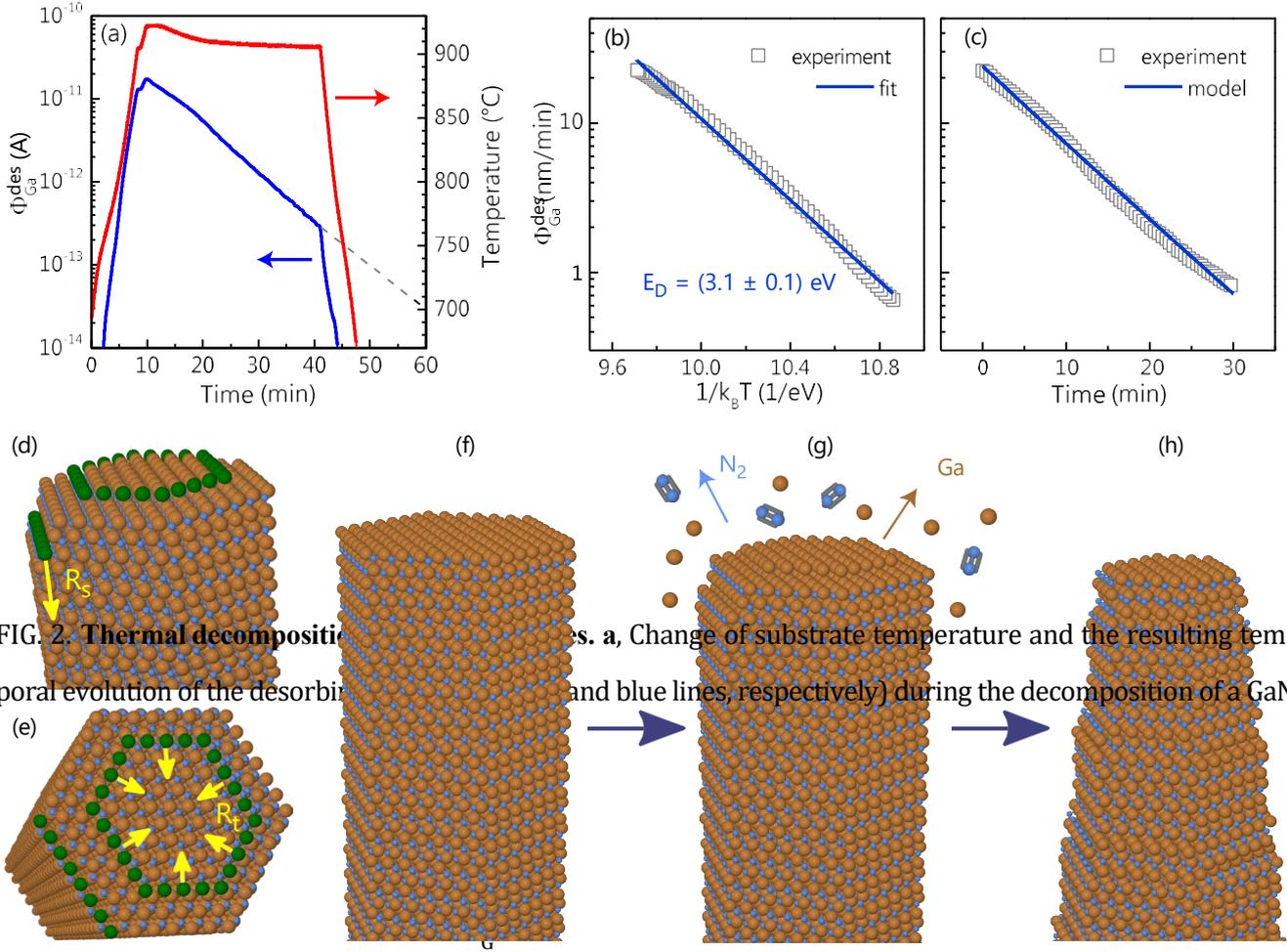

FIG. 2. **Thermal decomposition ... s. a,** Change of substrate temperature and the resulting temporal evolution of the desorbing ... (red and blue lines, respectively) during the decomposition of a GaN nanowire ensemble. The ion flux is given in units of ion current. The dashed line shows an exponential extrapolation of $\Phi_{Ga}^{des}$ to the background level. **b,** Arrhenius representation of the temperature dependence of $\Phi_{Ga}^{des}$ during the decomposition of a GaN nanowire ensemble. The line shows a fit with an activation energy of $E_D = (3.1 \pm 0.1)$ eV. **c,** Temporal evolution of $\Phi_{G}^{des}$ during thermal decomposition at 920 °C. The line shows the result of a fit using the modeling described in the Supplementary Information and yields layer-by-layer decomposition rates for the top facet and for the sidewalls of $R_s = 2.5 \times 10^{-4}$ and $R_t = 2.8 \times 10^{-3}$ s⁻¹, respectively. **d, e,** Schematic representation of the layer-by-layer decomposition process along the **d** sidewalls and **e** top facets. The green atoms denote the respective step edge and the yellow arrows indicate how the decomposition process at the side and top facets proceeds. **f–h,** Schematic representation of the morphology of a GaN nanowire during thermal decomposition in UHV. Panel **f** shows the initial morphology of the as-grown nanowire, **g** the morphology at the onset of lateral and axial decomposition, and **h** the final morphology, which is characterized by a pronounced tapering caused by the simultaneous decomposition of several atomic layers.

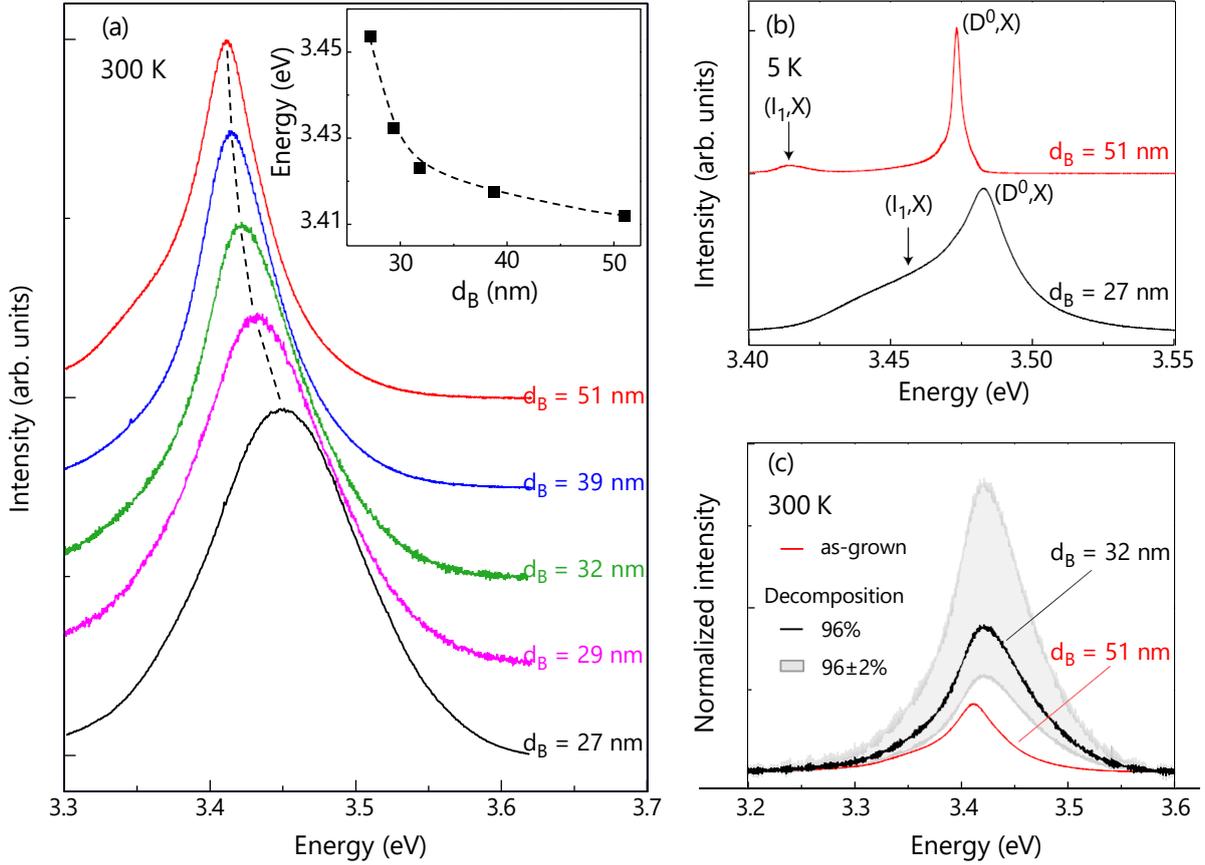

FIG. 3. **Dielectric confinement in ultrathin GaN nanowires. a**, Photoluminescence spectra at 300 K from an ensemble of as-grown GaN nanowires (top spectrum) and from partially decomposed GaN nanowire ensembles with different average base diameters $d_B$ as indicated in the figure. The spectra have been normalized and shifted vertically for clarity. The free exciton transition shows a 42 meV blueshift with decreasing diameter (see inset). The dashed lines are guides to the eye. This blueshift arises from the radial dielectric confinement of the exciton due to the dielectric mismatch between GaN and vacuum. **b**, At 5 K, the photoluminescence spectra are dominated by the recombination of excitons bound to donors. The weaker band highlighted by arrows arises from the recombination of excitons bound to basal plane stacking faults. Both excitonic transitions blueshift with decreasing $d_B$ due to dielectric confinement. **c**, Photoluminescence spectra recorded at 300 K from the ensemble of as-grown nanowires (red) and from the ensemble of thinned nanowires with $d_B = 32$ nm [Fig. 1(b)]. The black line shows the photoluminescence spectrum of the thinned nanowires after normalizing their emission intensity by the volume ratio. The grey shaded area shows the intensity uncertainty corresponding to the ±2% error from the QMS and RBS analysis.

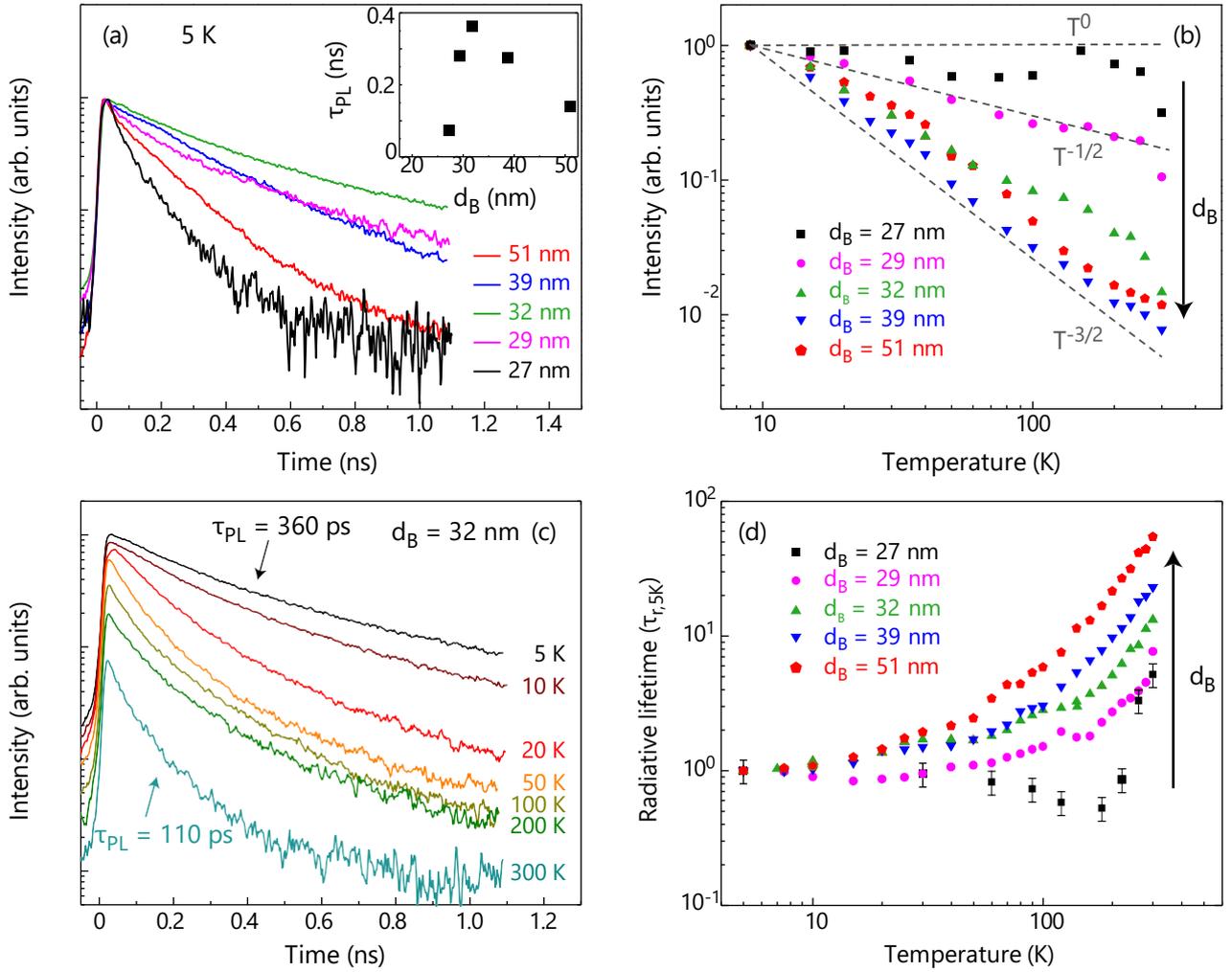

FIG. 4. **Radiative and nonradiative recombination of excitons in ultrathin GaN nanowires. a**, Photoluminescence transients recorded at 5 K for the same samples as in Fig. 3(a). The inset shows the photoluminescence decay time $\tau_{PL}$, obtained using an exponential fit, as a function of $d_B$. The nonmonotonic diameter dependence of $\tau_{PL}$ demonstrates that surface recombination at the nanowire sidewalls does not dominate recombination even for very thin nanowires. **b**, Spectrally integrated photoluminescence intensity as a function of temperature for the samples under investigation. The dashed lines are guides to the eye. The thinner the nanowires, the weaker the photoluminescence quenching with temperature and the larger the radiative efficiency at 300 K. **c**, Photoluminescence transients as a function of temperature for the ensemble of thinned nanowires with $d_B$ = 32 nm. The nonradiative lifetime only decreases by a factor of 3 over a temperature range from 5 to 300 K. This reduction is too small to account for the quenching of the photoluminescence intensity observed in panel **a** for this sample. **d**, Temperature dependence of the exciton radiative lifetime for the samples under investigation. The radiative lifetimes are given in units of the radiative lifetime at 5 K ($\tau_{r,5K}$). For all samples, the temperature dependence of the radiative lifetime is consistent with the one of the emission intensity. The thinner the nanowire, the weaker the increase in radiative lifetime with temperature and the larger the radiative e1ffi8ciency.